\documentclass[aps,pre,showpacs,onecolumn,superscriptaddress,floatfix]{revtex4}
\usepackage{graphicx,color,epsfig} \usepackage{color}

\usepackage{graphicx}
\usepackage{epsfig}
\usepackage{epstopdf}
\usepackage{color}
\usepackage{amsfonts}
\usepackage{amssymb,amsmath}
\usepackage{multirow}
\usepackage{pifont}
\newcommand{\R}{\color{red}} 
\newcommand{\G}{\color{green}}
\usepackage[utf8]{inputenc}
\usepackage{tikz}
\usepackage{multirow}
\newcommand{\cmark}{\ding{51}}%
\newcommand{\xmark}{\ding{55}}%

\begin{document}

\title{Chimera and anticoordination states in learning dynamics}

\author{Hayd\'ee Lugo}
 \affiliation{ICAE and Department of Economic Analysis. Universidad Complutense de Madrid, 28223 Madrid, Spain}
\bigskip
\author{ Juan Carlos Gonz\'alez-Avella}
 \affiliation{APSL S.L, Edifici Europa - Planta baja Galileo Galilei, Parc Bit 07121 Palma de Mallorca, Spain}
 \bigskip
\author{Maxi San Miguel}
 \affiliation{ IFISC (CSIC-UIB), Campus Universitat de les Illes Balears, 07122 Palma de Mallorca, Spain }

\begin{abstract}
In many real-life situations, individuals are dared to simultaneously achieve social
objectives of acceptance or approval and strategic objectives of coordination.
Since these two objectives may take place in different environments, a two-layer
network is the simple and natural framework for the study of such kind of
dynamical situations. In this paper we present a model in which the state of the agents
corresponds to one of two possible strategies.
They change their states by interaction with their
neighbors in the network. Inside each layer the agents interact by a social pressure
mechanism, while between the layers the agents interact via a coordination game.
From an evolutionary approach, we focus on the asymptotic solutions for all-to-all
interactions across and inside the layers and for any initial distribution of
strategies.
We find new asymptotic configurations which do not exist in a single isolated
social network analysis. We report the emergence and existence of chimera
states in which two different collective states coexist in the network. Namely, one layer
reaches a state of full coordination while the other remains in a dynamical state of
coexistence of strategies. In addition, the system may also reach a state of global
anticoordination where a full coordination is reached inside each layer but with
opposite strategies in each of the two network layers. We trace back the emergence of
chimera states and global anticoordination states to the agents inertia against
social pressure, referred here to as the level of skepticism, along with the degree of risk taken into account in a general coordination game.

\end{abstract}

\maketitle

\section{Introduction}

The coexistence of coherent and incoherent states has received much attention as an intriguing
manifestation of collective behaviour. This interesting  behaviour was first observed by Kuramoto et al. \cite{kb}
and then named it as chimera state \cite{as}. Although the literature about
chimera states started with
the study of interacting populations of oscillators in dynamical systems [\cite{mp}, \cite{rb}, \cite{so}, \cite{pa}, \cite{kbg}], it has been dizzily expanded to many fields in physics, chemistry, biology, etc.
Also in social systems, situations of two interacting populations in which one exhibits a
coherent or synchronized behaviour while the other is incoherent or
desynchronized are commonly observed. This phenomena has also been addressed from the conceptual framework of chimera states.

Models based on individual interactions have been introduced for opinion formation and
cultural dissemination  [ \cite{gac1} and \cite{gac2}]  in order to  analyze the emergence of localized coherent or chimeras states behavior in social contexts. The systems considered
consist in two populations of social agents mutually coupled through global interaction fields that account for the state of the majority of the agents in each population.
The internal interactions in each group have a condition that allow for non-interacting
states. Two examples of such dynamics have been analyzed: (i) Axelrod's model for social
influence \cite{a}, and (ii) a discrete version of a bounded confidence model for opinion formation \cite{dna}.
In both systems, there are localized coherent states for some parameter values, in which a group
reaches a homogenous or consensus state, while the other group remains in a disordered or
polarized state. In this paper we contribute further to the study of chimera states in social
systems, searching for them in the context of models of social coordination and learning dynamics \cite{redondo}.

Coordination is an important issue in economics studies, being studied within a Game Theory approach in many theoretical and 
experimental studies, [\cite{e}, \cite{vr}, \cite{gh}]. It has has become one of the most important challenges in modern societies: In 
spite of the fact that  individuals are now
more connected and handle more information due to technological progress, it seems that coordination to reach consensus is becoming an increasingly difficult goal.
It is common to observe how some societies become polarized either by ideological or political opinions, or,  the collective behaviour 
leads the population to states in which one part reaches a consensus while the other part behaves in an unstable manner, or it displays a dynamical 
coexistence of strategies.
It can be argued that such diversity of outcomes are the result of a constant search  for achieving simultaneously social and economic aims in modern societies.
On one side, individuals, influenced by others, seek for social acceptance and recognition
and on the other side they attempt to get higher gains.
Whenever the social and the economic concerns take place in two different social networks or environments, the population unavoidably splits 
into two disjoint target groups.
An example of different environments is the environments of family and friends versus the environments of work and business.

Here we present a simple model to illustrate such situation in which individuals of a population divided into two groups are dared to coordinate in
order to accomplish their social and economic goals.
Our framework for studying such kind of population is a two-layer network. Each layer corresponds to a group of the population and interactions within the group aim to satisfy social concerns, while interactions between agents of different layers aim to satisfy economic concerns.
In our system, the economic goal turns out to be reached when agents play the same strategy. Since there will be as many consensus as numbers of possible strategies a coordination problem arises. Also, a coordination problem to achieve social goals arises because
there is a skepticism by individuals to be influenced by social pressure.
A previous work of \cite{lsm} shows that for an initial
uniform distribution of two possible strategies the skepticism to follow the opposite strategy and the local connectivity are the driving forces  to accomplish
full coordination for this two-layer network.
Here, we consider the role of different initial conditions leading to different asymptotic states of the dynamics and we determine the basin of attraction of these states.
As one of our main results, we find two asymptotic states with nontrivial collective behaviour which can not be found in a single isolated network analysis.
A first outcome is a social analogue of a chimera state with coexistence of coherent and incoherent states.
In our model, one layer can reach a homogeneous state of full coordination while the other remains in a dynamical state of coexistence of strategies.
The second non-trivial asymptotic state is the anticoordination or polarization state in which the system reaches coordination with a different strategy in each layer.

The paper is organized as follows.
Section 2 introduces the general frame of our model. Based on a two-layer network, we describe the kind
of interactions inside and between layers and the dynamical rule for individuals to update their strategies.
Section 3 describes the possible asymptotic solutions of the collective dynamics reached by the system, as well as the basins of attraction to reach these solutions. 
 In particular we describe the nontrivial chimera and anticoordination states and their basins of attraction.
Section 4 shows bifurcation diagrams, in Sec. 4.1 for the case of a symmetric coordination game and in Sec. 4.2 for the case of an asymmetric coordination game.
Finally, Section 5 summarizes our conclusions. In the Appendix,
we present a mean-field theory for the time evolution of the system in the infinite size limit.

\begin{figure}[t]
  \begin{center}
  \includegraphics[width=0.6\linewidth,angle=0]{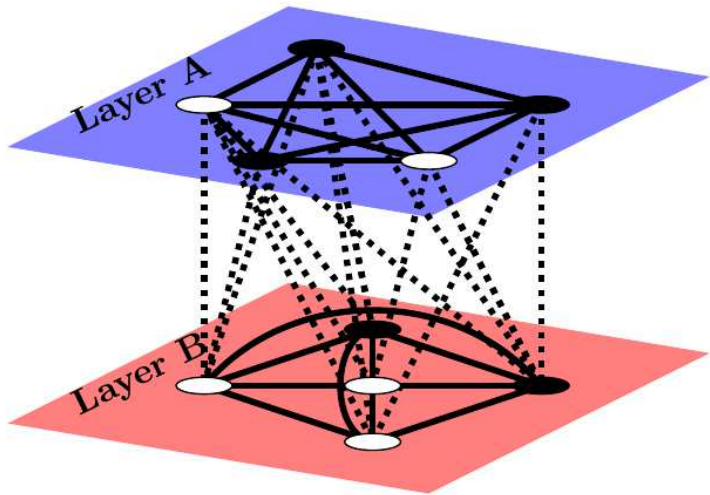}
  \end{center}
\caption{Sketch of a two-layer network. The nodes are connected to each other in a pairwise manner both inside the layers (intralayer links) and across the layers A and B  (interlayer links). Dotted lines describe the interlayer interactions and the solid lines describe the intralayer interactions. Black nodes stand for the agents playing strategy L, while white nodes stand for the agents playing strategy R in a coordination game.}
\label{multilayer}
  \end{figure}

We consider an individual based model consisting in a population in which individuals interact in two different groups, A and B of sizes $N_A$ 
and $N_B$ respectively. We take $N_A=N_B=N$.  Using the frame of a two-layer network (see Figure \ref{multilayer}), inside
each layer the individuals interact with social objectives and between layers they interact with strategic or
economic objectives,  as described in the following.

%Figura1
%  %%%%%%%%%%%%%%%%%%%%%%%%%% %%%%%%%%%%%%%%%%%%%%%%%%%%%%%%%%%%%%%%%%%%%%%%%%
% %%%%%%%%%%%%%%%%%%%%%%%%%%%%%%%%%%%%%%%%%%%%%%%%%%%%%%%%%%%%%%%%%%%%%%%%%%%%%%%

%%%%%%%%%%%%%%%%%%%%%%%%%%%%%%%%%%%%%%%%%%%%%%%%%%%%%%%%%%%%%%%%%%
 \subsection{Interactions between the layers: Coordination games}\label{CG}

In Game Theory [\cite{e}, \cite{vr}, \cite{gh}], the coordination game is a prototype model of strategic game with multiple Nash equilibria. Each player can choose a strategy, for example driving on the left ($L$) or on the right ($R$) side. In equilibrium none of the players
want to change their choice of strategies, given the strategies of the other players.
In this game  players share the goal of coordinating on the same action, however, when they choose their strategies simultaneously, without knowing what the other players will choose, it becomes highly difficult for them to coordinate in the desirable way.

The coordination dynamics proceeds as follows. In each time step, each agent in a layer plays with each agent in the other layer a pairwise coordination game.
Table  \ref{gcg} shows the pay-off normal form representation of a two-person, two-strategy coordination
game: For example if one agent plays $L$ and the other plays $R$, the pay off for the former is $0$, and $-b$ for the latter.
We focus our analysis on two parametric
settings, a pure or symmetric coordination game (SCG) in which $s = 0$ and $b=0$ and
a general or asymmetric coordination game (ACG) in which $s=1$ and $b>0$.
The profiles $(L,L)$ and $(R,R)$ are the two pure strategy Nash equilibria
in both settings. A problem of equilibrium selection is present in both settings. In the symmetric coordination game both equilibria are equivalent. However, in the general coordination game the higher
payoff is achieved when agents play the profile  $(R,R)$,  called  the Pareto (payoff) dominant equilibrium.
However, for $b >1$, the profile $(L,L)$ becomes the risk
dominant equilibrium,  in the sense of \cite{hs}, since agents risk less by coordinating on $(L,L)$. Then, the
socially efficient solution $(R,R)$  turns risky so that the possibility
of coordinating in it decreases as $b$ increases. The parameter $b$ becomes a measure of risk for playing the strategy $R$. For a complete review, see \cite{w}.

% Table I
\begin{table}[]
\begin{center}
\begin{tikzpicture}[scale=2.1]
\draw (1,0) grid (3,2);
\node at (0.5,1.5) {$ L$};
\node at (0.5,0.5) {$R$};
\node at (1.5,1.5) {$1, 1$};
\node at (2.5,1.5) {$0, -b$};
\node at (1.5,0.5) {$-b, 0$};
\node at (2.5,0.5) {$1 + s, 1 + s$};
\node at (1.5,2.2) {$L$};
\node at (2.5,2.2) {$R$};
\end{tikzpicture}
\end{center}
\caption{Payoff matrix for a two-person, two-strategy coordination game}
\label{gcg}
\end{table}%
%
%%%%%%%%%%%%%%%%%%%%%%%%%%%%%%%%%%%%%%%%%%%%%%%%%%%%%%%%%%%%%%%%%%%%%%%%%%%%

\subsection{Interactions inside the layers: the effect of social pressure}
Inside each layer,
searching for social acceptance and approval, each agent observes the strategies being played by her partners.
An agent may not feel at ease with her strategy when such strategy is not as popular as she wants in her social environment.
The level of skepticism in the population is calibrated by a threshold $T$ that determines the effect of the social pressure exerted on an individual.
 The criterion used by each player $i$  is to measure how well she is doing by comparing the share  of
agents who are playing the opposite strategy to hers,  denoted by $d_i $, with the threshold
 $T \in[0,1]$.
We may distinguish two types of populations. {\it Herding population}, for $T<0.5$,
in which
 individuals are influenced by low levels of popularity of the opposite
 strategy, and, {\it skeptical population}, for $T
 > 0.5$,  where
the social pressure has a weak effect on individuals: a feeling of disapproval only arises for high levels of popularity of the opposite strategy.
Therefore,
when $d_i >T$,   the social pressure is effective  because player $i$ generates a feeling of non-acceptance about the strategy
 she is currently playing and she is willing to revise it.
%
%%%%%%%%%%%%%%%%%%%%%%%%%%%%%%%%%%%%%%%%%%%%%%%%%%%%%%%%%%%%%%%%%%%%%%%%%%%%%%
\subsection{Inter-layer and intra-layer objectives: The degree of satisfaction}
In the interactions across and inside the layers agents intend to satisfy two different objectives:
social objectives of acceptance and approval, and strategic objectives of coordination. These objectives give rise to
 two sources of satisfaction:
strategic satisfaction in terms of the payoff obtained in the coordination game
and social satisfaction in terms of the effect of social pressure.
Therefore, there are four degrees of satisfaction described in Table \ref{sa}, where
$\pi_i$ is the aggregate payoff that agent $i$ in a layer gets playing with all the other agents in the other layer.
%Table II
\begin{table}[]
\begin{center}
\begin{tabular}{c|c|c|c}
%\hline
 S  &  P1 & P2  &	U\\ \hline
 $\pi_i=(1+s) N$  & $\pi_i=(1+s) N$ & $\pi_i < (1+s)  N$  &$\pi_i< (1+s) N$	\\
$d_i < T$ &  $d_i > T$ & $d_i < T$ & $d_i > T$ \\  \hline
\end{tabular}
\caption{Degrees of dissatisfaction according to the fulfilment of social and strategical objectives.}
\label{sa}
\end{center}
\end{table}%

The value of $s$ is derived from the parametric setting of
the coordination game.  When $s=0$ the equality
$\pi_i=N$ shows that the player $i$ coordinates with all the members of the other layer
in the symmetric game. Then we say that
agent $i$ is strategically satisfied. In the case of a general
coordination game, $s=1$,  an agent is strategically satisfied
when the coordination is  on the socially
efficient solution, {\it i.e.} the Pareto dominant strategy.  When
$d_i < T$ the share of agents inside the layer of player $i$ who play
the same strategy as she does is high enough so the player $i$ feels
socially satisfied with her current strategy. Then, the level of
satisfaction of  agent $i$ can be: S (satisfied) when she is both
socially ($d_i < T$) and strategically ($\pi_i=(1+s) N$) satisfied,
P1 or P2 (partially satisfied) when she is either socially ($d_i<T$)
or strategically ($\pi_i = (1+s) N$) satisfied and is U
(unsatisfied) when she is both socially ($d_i >T$) and strategically
($\pi_i < (1+s) N$) unsatisfied.
%
%%%%%%%%%%%%%%%%%%%%%%%%%%%%%%%%%%%%%%%%%%%%%%%%%%%%%%%%%%%%%%%%%%%%%%%%%%%
\subsection{Learning dynamics}
The learning dynamics in the system is described by the update rule used to change strategy:
Following the learning dynamics of \cite{lsm} and \cite{glsm}, at each elementary time step each player plays the coordination game
with all the members in the other layer. Once the game is over
and a payoff is assigned to each player, each agent observes and measures the popularity
of her strategy in her own group. As a result, a level of satisfaction arises.
Then, she might change her strategy impelled by her
level of satisfaction. The process is repeated setting payoffs to
zero.
The synchronous update rule in which each player can change her
current strategy according to her level of satisfaction is described
as follows,
\begin{enumerate}
\item If her level of satisfaction is S, she remains with the same
  strategy.
\item If her level of satisfaction is P1 or P2, she imitates the
  strategy of her best performing agent inside the layer
  in case that such agent has received a larger payoff than the player
  herself, otherwise she remains with the same strategy.
\item If her level of satisfaction is U, she changes her current
  strategy.
\end{enumerate}

Although the update rule takes place inside the layers,
individuals change their strategies by both
social and strategic considerations.  The imitation of the
best performing individuals in her social environment aims
to capture the individual behaviour observed in many complex real life situations.
%

%%%%%%%%%%%%%%%%%%%%%%%%%%%%%%%%%%%%%%%%%%%%%%%%%%%%%%%%%%%%%%%%%%%%
\section{ Results}

\subsection{Asymptotic solutions}

Analytical equations for our model and their asymptotic solutions are discussed in the Appendix.
For the general case of the asymmetric coordination game these solutions, described below, are the following:

{\bf Solution I:} Coordination in strategy $L$: All agents in both layer play strategy $L$. It is linearly stable and exists for any $T \in [0,1]$.

{\bf Solution II:} Coordination in strategy $R$: All agents in both layer play strategy $R$. It is linearly stable and exists for any $T \in [0,1]$.

{\bf Solutions III:} Coexistence of strategies. 
These solution exist for any $T \ne 1$ and it
occurs in two ways:

(1)  unstable fixed points and,

(2)  family of marginally stable periodic solutions.

{\bf Solutions III-a and III-b: Chimera solutions.}
This is an interesting case of coexistence of two distinct solutions, namely, solutions I and III.
One of the layers goes to the absorbing state
of coordination in strategy $L$, namely layer A for Solution III-a and layer B for solution III-b, while the other layer goes to a
dynamical state of coexistence of strategies, solution III type (2). We also find the case of coexistence of two solutions in which one layer coordinates in $L$ and the other layer remains disordered with both strategies coexisting in the same proportion (Solution III type (1)). These solutions only appear when agents are playing the asymmetric coordination game. They exist for almost any $T < 0.5$ and almost any $b > 0$.

Although the strategies $L$ and $R$ are not equivalent in the asymmetric coordination game because the first is the socially undesired and the second the socially desired outcome, solutions III-a and III-b are equivalent in the sense that the layer reaching the absorbing state $L$ is determined by the initial conditions of strategies in the two layers.
We refer to these solutions as chimera states because of the coexistence of an ordered layer and a disordered layer. The disordered layer can be in a dynamical state (solution III (2)) or in a configuration in which the number of agents playing each strategy is equal and constant in time (solution III (1)).

{\bf Solutions IV} Anticoordination states, layer A coordinates in strategy $L$ while layer B in strategy $R$.

{\bf Solutions V} Anticoordination states, layer A coordinates in strategy $R$ while layer B in strategy $L$.

Solutions IV and V exist and they are linearly unstable and exist for any $T \in [0.5,1]$.

\begin{figure}[t]
 \begin{center}
    \includegraphics[width=0.6\linewidth,angle=0]{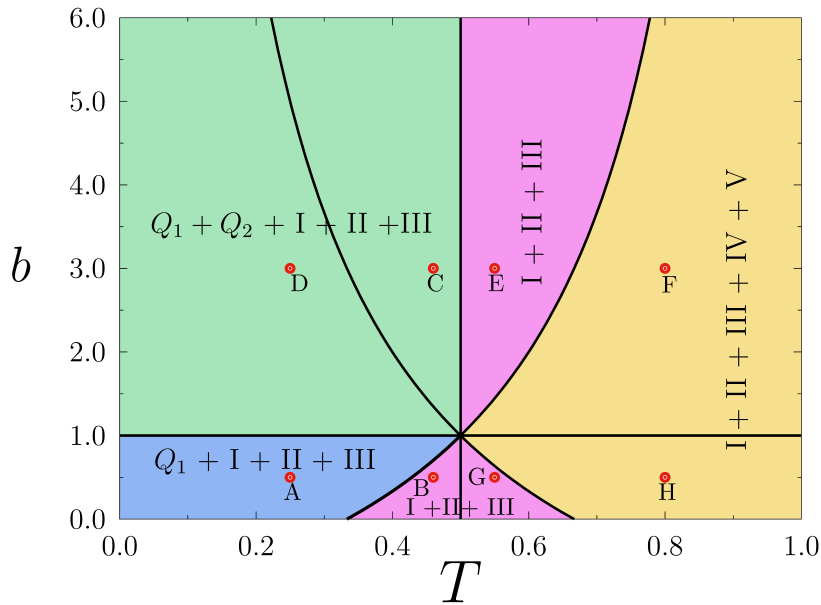}
 \end{center}
 \caption{Domain of existence of the different asymptotic solutions in the Asymmetric Coordination Game (ACG) as a function of the
threshold $T$ and the risk parameter $b$ as obtained from the analytic solution of the equations discussed in the Appendix. The two curves ( $T = \frac{1+b}{3+b}$ and
 $1 -T = \frac{1+b}{3+b}$) and the two lines ($b=1$, $T= 0.5$) divide the figure in eight zones from (A) to (H). In each of these zones the asymptotic solutions that exist are indicated. The red dots correspond to values of $T$ and $b$ used in Fig. 5.}
 \label{F2}
 \end{figure}

We summarize in Figure \ref{F2} the domain of existence of the different asymptotic solutions for the general case of the asymmetric coordination game as a function of the threshold $T$
and the risk $b$ parameters. For this general case the parameters of the pay-off matrix take the values $s=1$ and $b>0$ for any $T \in [0,1]$. Asymptotic solutions of the dynamics depend on initial conditions (see below Fig. 5), and in this sense we refer to $Q_1$ as those chimera states, solutions III-a and III-b, that are reached from initial conditions such that ${x_{aa}}^0 + {x_{bb}}^0 <1$, where ${x_{aa}}^0$ and ${x_{bb}}^0$ are the initial conditions for $x_{aa}$ and $x_{bb}$ respectively. Likewise we refer to $Q_2$ as those chimera states, solutions III-a and III-b, that are reached from initial conditions such that ${x_{aa}}^0 + {x_{bb}}^0 >1$. It turns out that solutions $Q_1$ exist for $b>0$,  zones A, C and D in Figure \ref{F2}, while $Q_2$ only exist for $b>1$, zones D and C. The range of values of $b$ and $T$ in which the system reaches solutions $Q_1$, $Q_2$, I ,II and III corresponds to zones C and D in the Figure. The difference between zones C and D refers to the areas of the basins of attraction of each solution, as explained in Sect. 3.2.
For the case of symmetric coordination game,  the parameters take the values
$s=0$ and $b=0$. The asymptotic solutions in this case are the same as those for the asymmetric coordination game
except for Solutions III-a and III-b, the Chimera states. Chimera states could not be found for any level of skepticism in the case of the symmetric coordination game.

%%%%%%%%%%%%%%%%%%%%%%%%%%%%%%%%%%%%%%%%%%%%%%%%%%%%%%%%%%%%%%%%%%%%%%%%%%
 % Figure 2

%%%%%%%%%%%%%%%%%%%%%%%%%%%%%%%%%%%%%%%%%%%%%%%%%%%%%%%%%%%%%%%%%%%%%%%%%%%%%

%%%%%%%%%%%%%%%%%%%%%%%%%%%%%%%%%%%%%%%%%%%%%%%%%%%%%%%%%%%%%%%%%%%%%%%%%%%%%

Given that the agent population is divided in two layers, A and B,
we define $x_{aa}$ as the fraction of individuals playing strategy $R$ in
layer $A$, and $x_{bb}$ as the fraction of individuals playing strategy $R$ in layer $B$. To describe the different asymptotic solutions we also introduce the order parameter $n_{AB}$ giving the density of inter-layer active links, i.e.,
the proportion of links connecting agents in different layers with opposite strategies.
The order parameter $n_{AB}$  can be written in terms of  $x_{aa}$ and $x_{bb}$ by,
\[
n_{AB} = x_{aa}( 1 - x_{bb}) + x_{bb} (1- x_{aa})
\]

The different solutions  described in terms of the  asymptotic values  of  $n_{AB}$,  $x_{aa}$ and $ x_{bb}$ are shown in Table \ref{as}.
These solutions follow from the mean-field analysis described in the Appendix. We next describe these solutions as obtained from numerical simulations.
In these simulations, we have fixed the system size $N_A + N_B= 2000$ where $N_A=N_B=1000$.

%table 3
\begin{table}[]
\begin{tabular}{|l|l|l|l|l|l|l|}
\hline
Asymptotic state                           & $n_{AB}$                                              & $x_{aa}$    & $x_{bb}$          & solution          & SCG & ACG \\ \hline
\multirow{2}{*}{Coordination}         & \multicolumn{1}{c|}{\multirow{2}{*}{0}} & 0                 & 0                 & I                      & {\G \cmark }  & {\G \cmark }     \\ \cline{3-7}
                                                      & \multicolumn{1}{c|}{}                            & 1                 & 1                 & II                     &  {\G \cmark }    &   {\G \cmark }   \\ \hline
\multirow{2}{*}{Anticoordination}   & \multicolumn{1}{c|}{\multirow{2}{*}{1}} & 0                 & 1                 & IV                     &  {\G \cmark }    &  {\G \cmark }    \\ \cline{3-7}
                                                      & \multicolumn{1}{c|}{}                            & 1                 & 0                 & V                      &   {\G \cmark }   &  {\G \cmark }    \\ \hline
\multirow{3}{*}{Coexistence of strategies} & $\frac{1}{2}$                            & $\frac{1}{2}$ & $\frac{1}{2}$ & \multirow{3}{*}{III}   & {\G \cmark }     &  {\G \cmark }  \\ \cline{2-4} \cline{6-7}
                                                      & $\frac{1+b}{3+b}$                               & $\frac{1+b}{3+b}$ & $\frac{1+b}{3+b}$ &                 &  {\R \xmark}   &  {\G \cmark }   \\ \cline{2-4} \cline{6-7}
                                                       & u+v -2uv                                             & u, 1-u            & v, 1-v            &            &   {\G \cmark }   & {\G \cmark }     \\ \hline
\multirow{4}{*}{Chimera states}                 &  v,1-v                                                  & 0                   & v, 1-v             & \multirow{2}{*}{III-a} &   {\R \xmark}     &  {\G \cmark }     \\ \cline{2-4} \cline{6-7}
                                                       &  $\frac{1}{2}$                                     & 0                   & $\frac{1}{2}$   &                        &  {\R \xmark}      &    {\G \cmark }   \\ \cline{2-7}
                                                       &        u, 1-u                                          & u, 1-u               & 0             & \multirow{2}{*}{III-b} &   {\R \xmark}     &   {\G \cmark }   \\ \cline{2-4} \cline{6-7}
                                                       &  $\frac{1}{2}$                                      &    $\frac{1}{2}$           & 0             &                        &   {\R \xmark}     &    {\G \cmark }   \\ \hline
\end{tabular}
\caption{Asymptotic states and associated values of $x_{aa}, x_{bb}$ and $n_{AB}$.  The last two columns indicate existence or non-existence of the state in the Symmetric Coordination Game (SCG) and in the Asymmetric Coordination Game (ACG). The parameter $b$ corresponds to the risk parameter of the ACG and the numbers $u,1-u$ and $v, 1-v$  for $u, v \in (0,1)$ describe the family of periodic solutions in layer A and layer B respectively. Chimera states in which the three variables $x_{aa}, x_{bb}$ and $n_{AB}$ take constant values, either 0 or 1/2, correspond to the case in which the disordered layer is in a solution III (1).}
\label{as}
\end{table}

\begin{figure}[]
\begin{center}
 \includegraphics[width=1\linewidth,angle=0]{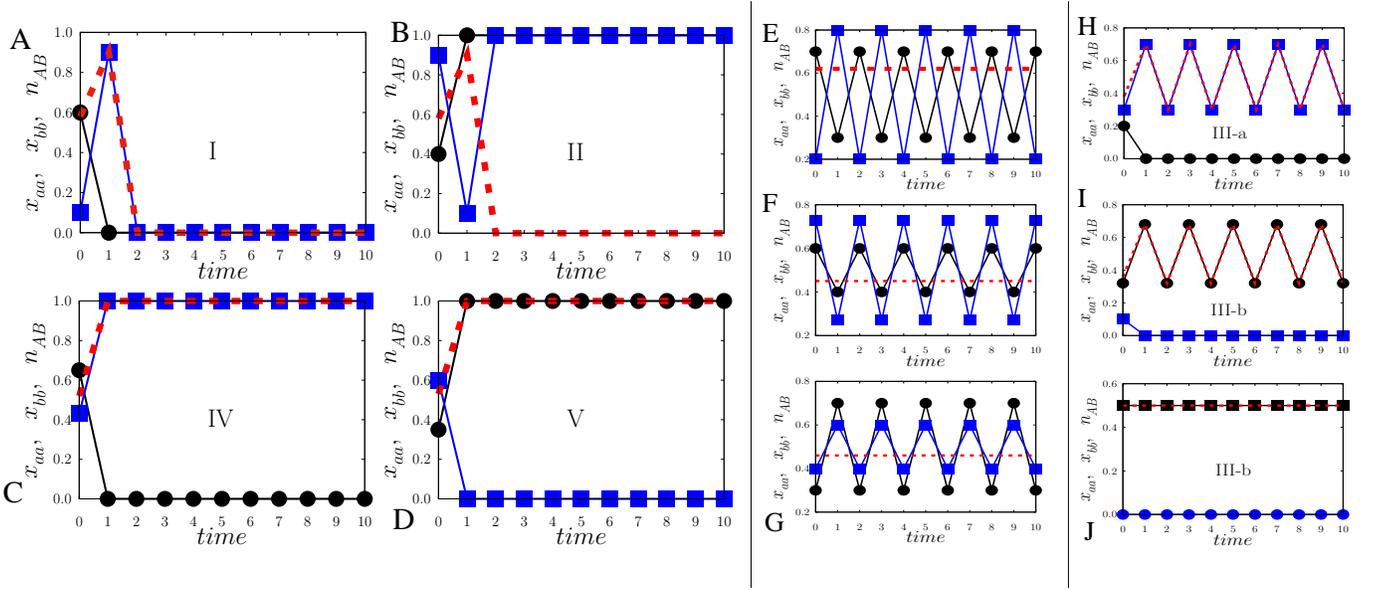}
\end{center}
 \caption{ Time evolution of $x_{aa}$ (circles), $x_{bb}$ (squares) and
the density of active inter-layer links $n_{_{AB}}$ (dashed lines). Panels A to D show the time evolution to coordination (solutions I and II) and anticoordination states (solutions IV and V) for symmetric game ($s=0, b=0$) with $T=0.75$.
Panels E, F and G show the phase and anti-phase oscillations (solutions III (2))
of the fraction of strategies  $x_{aa}$ and $x_{bb}$  for symmetric game ($s=0, b=0$) with $T=0.25$.
Panels H, I and J show the temporal evolution of $x_{aa}$ and $x_{bb}$ and the active inter-layer links $n_{_{AB}}$,
 for the asymmetric coordination game ($s=1, b=0.5$)  with  $T=0.25$.
Panels H and I show the chimera states, solution III-a and III-b respectively and
Panel J shows the case when the initial and the final states are the same when $x_{aa} = 0$ and $x_{bb}= 0.5$.}
  \label{F3}
 \end{figure}

Figure \ref{F3} shows the time evolution of $x_{aa}$ and $x_{bb}$, as obtained from the simulation of our individual based model, for different initial conditions that lead to the asymptotic solutions  I, II, III, IV and V.
Figures \ref{F3}A  to  \ref{F3}G correspond to  the symmetric coordination game (SCG), while chimeras states appearing in the asymmetric coordination
game (ACG) are shown in Figures \ref{F3}H to \ref{F3}J.
In Figures \ref{F3}A and \ref{F3}B the order parameter $n_{{AB}}=0$,
indicates that the system goes to an absorbing state in which the
agents in both layers play the same strategy.
Figure \ref{F3}A
shows that  after a short transient time the fractions $x_{aa} = x_{bb} = 0$
and the density of inter-layer active links is $n_{AB}=0$
corresponding to solution I,
while for Figure \ref{F3}B after a short transient time the fractions $x_{aa} = x_{bb} = 1$
and $n_{AB}=0$
corresponding to solution II.

In Figures \ref{F3}C (solution IV) and \ref{F3}D (solution V), the
value $n_{{AB}} = 1$ indicates that in both
cases the system goes to an anticoordination absorbing state, i.e.
agents in each layer are playing opposite strategies.  This state of anticoordination
can emerge  only in skeptical populations where $T > 0.5$.  It is interesting to notice that the layer with an initial higher proportion of $R$
is the layer that ends playing $L$.
A complete analysis of the anticoordination solutions for a skeptical two-group population can be found  in \cite{glsm}.

%%%%%%%%%%%%%%%%%%%%%%%%%%%%%%%%%%%%%%%%%%%%%%%%%%%%%%%%%%%%%%%%%%%%%%%%%%%%%%%%%%%%
%Figura 3
 %%%%%%%%%%%%%%%%%%%%%%%%%%%%%%%%%%%%%%%%%%%%%%%%%%%%%%%%%%%%%%%%%%%%%%%%%%%%%%%%%%%%
%Figura 4
\begin{figure}[]
 \begin{center}
 \includegraphics[width=1\linewidth,angle=0]{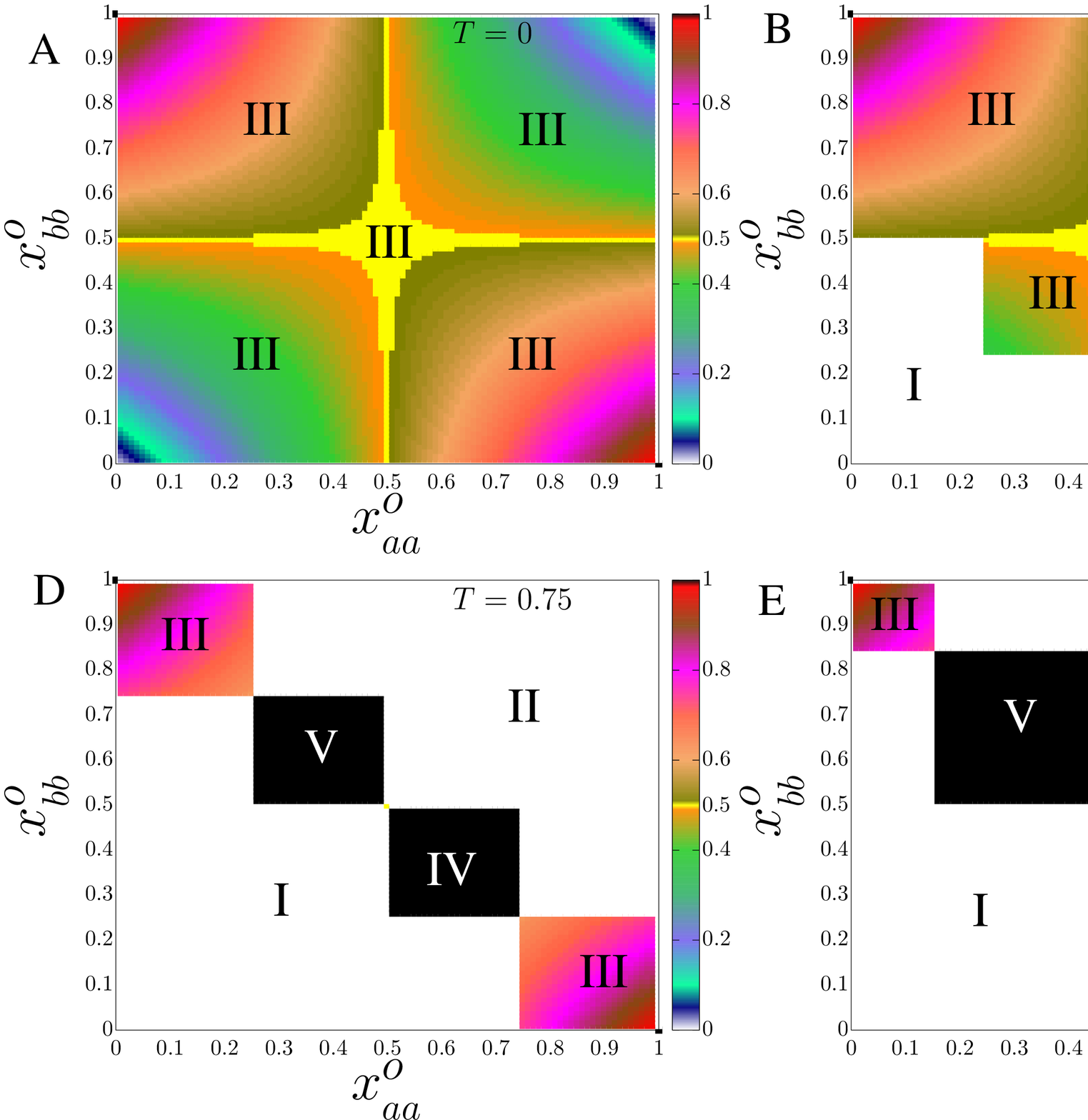}
 \end{center}
 \caption{Plot in color scale of the fraction of active links between layer A and B in the asymptotic solution of the dynamics as a function of
   the initial density of $x_{aa}$ and $x_{bb}$ for the Symmetric Coordination Game ($s=0$,$b=0$). The color scale defines the values
   of the fraction of actives links, $n_{AB}=1$ black color and $n_{AB}=0$
   white color. Asymptotic solutions are as indicated. Panels A to H correspond to different values
   of $T$:A) $T=0$, B) $T=0.25$, C) $T=0.5$, D) $T=0.75$, E) $T=0.85$ and F) $T=1$. System size, $N_A+N_B=1000+1000=2000$. }
  \label{F4}
 \end{figure}
 %%%%%%%%%%%%%%%%%%%%%%%%%%%%%%%%%%%%%%%%%%%%%%%%%%%%%%%%%%%%%%%%%%%%%%%%%%%%%%%%%%%%
 %Figura 5
  \begin{figure}[]
 \begin{center}
 \includegraphics[width=1\linewidth,angle=0]{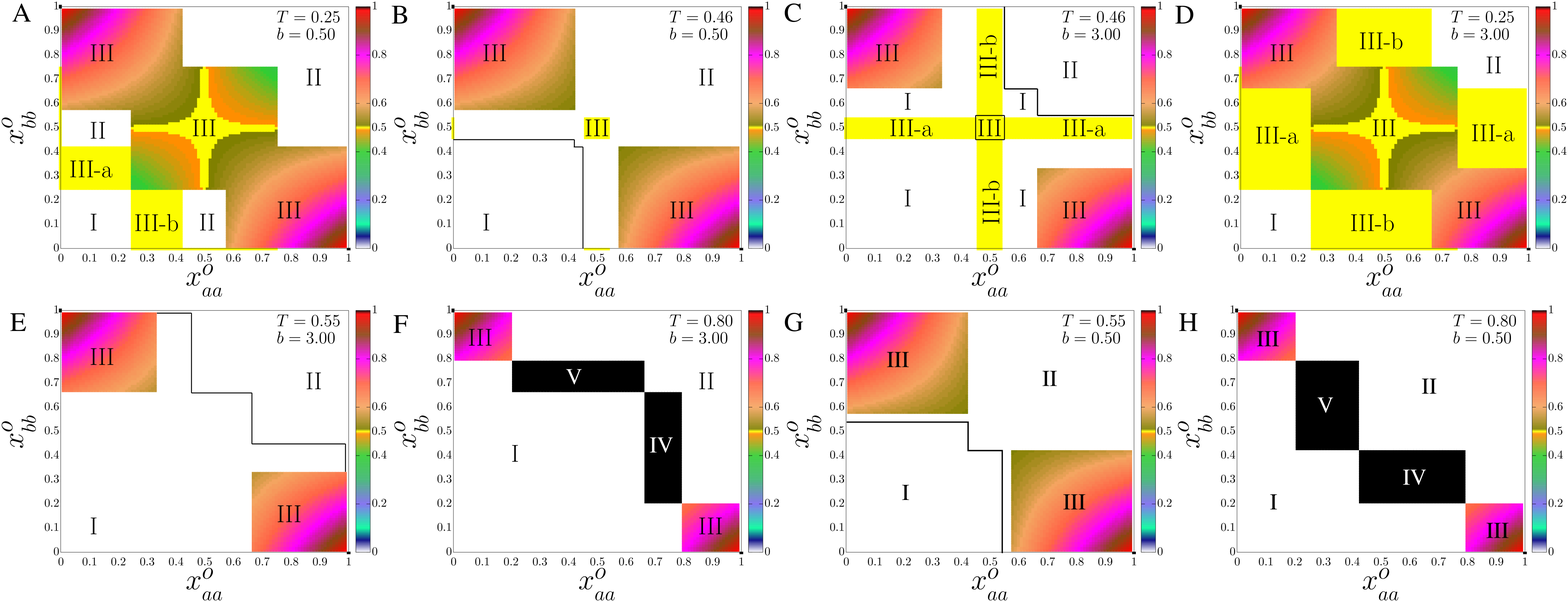}
 \end{center}
 \caption{Plot in color scale of the fraction of active links between layer A and B in the asymptotic solution of the dynamics as a function of
   the initial density of $x_{aa}$ and $x_{bb}$ for the Asymmetric Coordination Game. The color scale defines the values
   of the fraction of actives links, $n_{AB}=1$ black color and $n_{AB}=0$
   white color. Asymptotic solutions are as indicated. Panels A to H correspond to the values of the threshold $T$ and parameter $b$ indicated by red dots in Fig. 2. A) $T=0.25, b=0.50$, B) $T=0.46, b=0.50$, C) $T=0.25, b=0.30$, D) $T=0.25, b=3.00$, E) $T=0.55, b=3.00$, F) $T=0.80, b=3.00$, G) $T=0.55, b=0.50$, H) $T=0.80, b=0.50$.
   System size, $N_A+N_B=1000+1000=2000$.}
  \label{F5}
 \end{figure}
 %%%%%%%%%%%%%%%%%%%%%%%%%%%%%%%%%%%%%%%%%%%%%%%%%%%%%%%%%%%%%%%%%%%%%%%%%%%%%%%%%%%%

It is important to note that there exist absorbing states other than solutions  I, II, IV and V. They correspond to
an unstable fixed point $x_{aa} = x_{bb}= r$ of the dynamics for $0< r <1$. These solution correspond to the classification
(1) of Solution III, 
namely,  $r=1/2$ in the SCG for all $T \in (0,1)$ and  in the ACG for $T< 1/2$, or the fixed point  $r= \frac{1+b}{3+b} $ in  ACG when $T>1/2$.

Figures \ref{F3}E to \ref{F3}G display the temporal evolution of the system for marginally stable
periodic solutions (solutions III (2)) in the case of the symmetric coordination game.
The asymptotic dynamical configurations
show phase (Figure \ref{F3}F and \ref{F3}G) and anti-phase (Figure \ref{F3}E) oscillations of strategies between the two layers.
Note that $0 < n_{{AB}} < 1$ remains constant during these oscillations.

The Chimera solutions are illustrated in Figure  \ref{F3}H to \ref{F3}J
for the ACG with parameter values $T=0.25$, $b=0.5$.
Figure \ref{F3}H corresponds to a solution III-a, in which
all agents play strategy $L$ in layer A, i.e $x_{aa}=0$
while a configuration of dynamical coexistence of strategies takes
place in layer B, i.e. $0<x_{bb}<1$.
Note that the fraction of agents that choose to play strategy L or R
in layer B changes over time, so that we also observe an
oscillation of $n_{_{AB}}$.
Figure \ref{F3}I corresponds to a solution III-b. In this
case the behaviour is similar to the one in Figure \ref{F3}H, but now the $L$
coordinating absorbing state occurs in layer B, while the
dynamical coexistence of strategies appears in layer A.
We also notice that possibly similar solutions with $x_{aa} \ne 0$ constant, and $x_{bb}$
oscillating are not found in our model due to the nonequivalence of the $L$ and $R$ strategies in the ACG. In Figure \ref{F3}J we illustrate the particular case of a chimera state in which one layer coordinates in $L$, in this case layer B, while the other layer remains in a solution III (1) in which agents in that layer start and continue playing for all times both strategies with equal proportions.
The particular collective behaviors described by solutions III-a and III-b,
are the social analogue of a chimera state
arising in two interacting populations of oscillators observed in
dynamical systems.  In general a chimera state describes a situation where two
populations that interact with each other, one exhibits a
coherent or synchronized behaviour while the other is incoherent or
desynchronized.  Likewise we have two populations of interacting agents such that one reaches an absorbing state, while the other remains in a dynamically disordered state. Our chimera states only arise in our system when the population is {\it herding} ($T< 0.5$)
and play an asymmetric coordination game.
This means that beside the initial distribution of strategies, a herding behaviour is the mechanism that allows to reach
chimera states when the two equilibria of the coordination game are not equivalent in terms of payoff.

%%%%%%%%%%%%%%%%%%%%%%%%%%%%%%%%%%%%%%%%%%%%%%%%%%%%%%%%%%%%%%%%%%%%%%%%%%%%%%%%%%
\subsection{Basins of attraction: The global picture}
Depending on the initial conditions for
$x_{aa}$ and $x_{bb}$, the system
reaches different asymptotic solutions characterized by their value of the order
parameter $n_{AB}$.
Extensive numerical simulations of our individual based model are summarized in Figure \ref{F4} and Figure \ref{F5} that show the basins of attraction of the different asymptotic solutions in terms of the
initial densities of $x_{aa}$ and $x_{bb}$, and for different values of the threshold parameter $T$ for the SCG  and for different values of $T$ and $b$ for the ACG, respectively.
The color code defines the solutions in terms of the fraction of inter-layer active links.
Both figures show how the basins of attraction in terms of the initial conditions are determined by the value of the threshold parameter $T$ in the case of the SCG and the threshold $T$ along with the parameter $b$ in the case of the ACG.
 %%%%%%%%%%%%%%%%%%%%%%%%%%%%%%%%%%%%%%%%%%%%%%%%%%%
%
\section{ Bifurcation diagrams }
In this section we consider possible transitions among the different solutions discussed before. These transitions are described by means of bifurcations diagrams obtained in terms of the control parameters $T$ and $b$.

\subsection{Symmetric Coordination Game}
\begin{figure}[]
 \begin{center}
  \includegraphics[width=1\linewidth,angle=0]{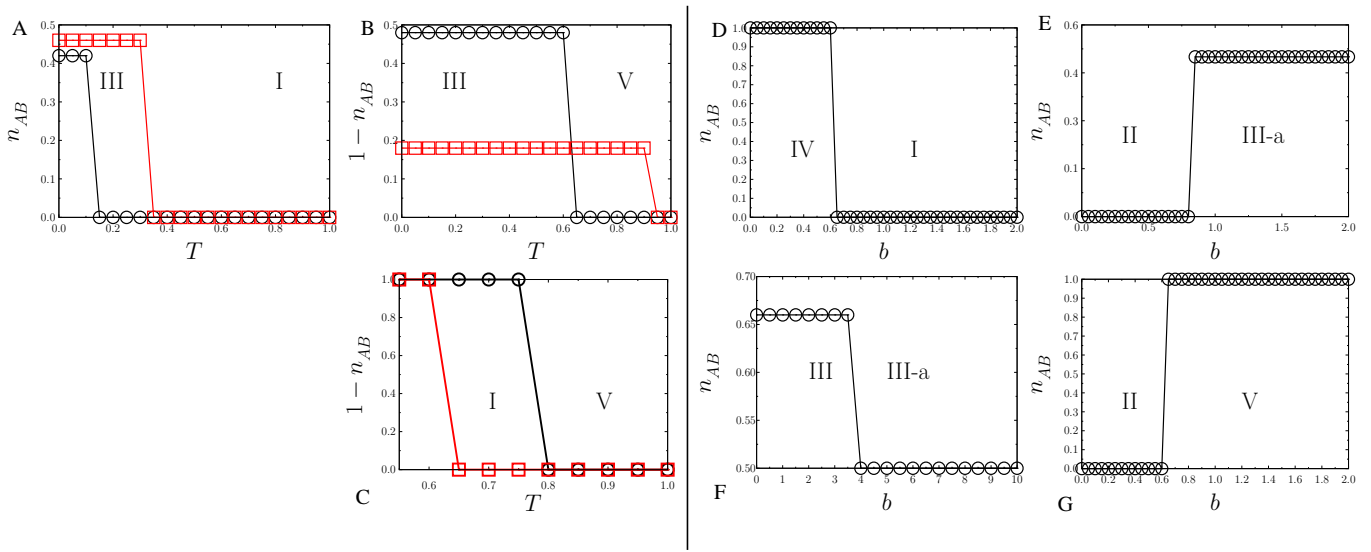}
 \end{center}
 \caption{Bifurcation diagrams for the average of $n_{AB}$ as function of $T$ for fixed initial
 conditions for the SCG, panels (A-C), and for the ACG, panels (D-G).
 Panel A: $x_{aa} = 0.40$  and $x_{bb} = 0.10$ (circles), $x_{aa} = 0,40$ and $x_{bb} = 0.30$ (squares).
 Panel B: $x_{aa} = 0.40$  and $x_{bb} = 0.10$ (circles), $x_{aa} = 0,10$ and $x_{bb} = 0.90$ (squares).
Panel C: $x_{aa} = 0.25$  and $x_{bb} = 0.51$ (circles), $x_{aa} = 0,37$ and $x_{bb} = 0.52$ (squares).
Panel D: $x_{aa} = 0.30$  and $x_{bb} = 0.45$  with $T = 0.75$.
Panel E: $x_{aa} = 0.10$  and $x_{bb} = 0.48$  with $T = 0.15$.
Panel F: $x_{aa} = 0.25$  and $x_{bb} = 0.1$   with $T = 0.7$.
Panel G: $x_{aa} = 0.45$  and $x_{bb} = 0.65$  with $T = 0.75$.
}
  \label{F6}
 \end{figure}

\begin{figure}[h]
\begin{center}
%A
 \includegraphics[width=1\linewidth,angle=0]{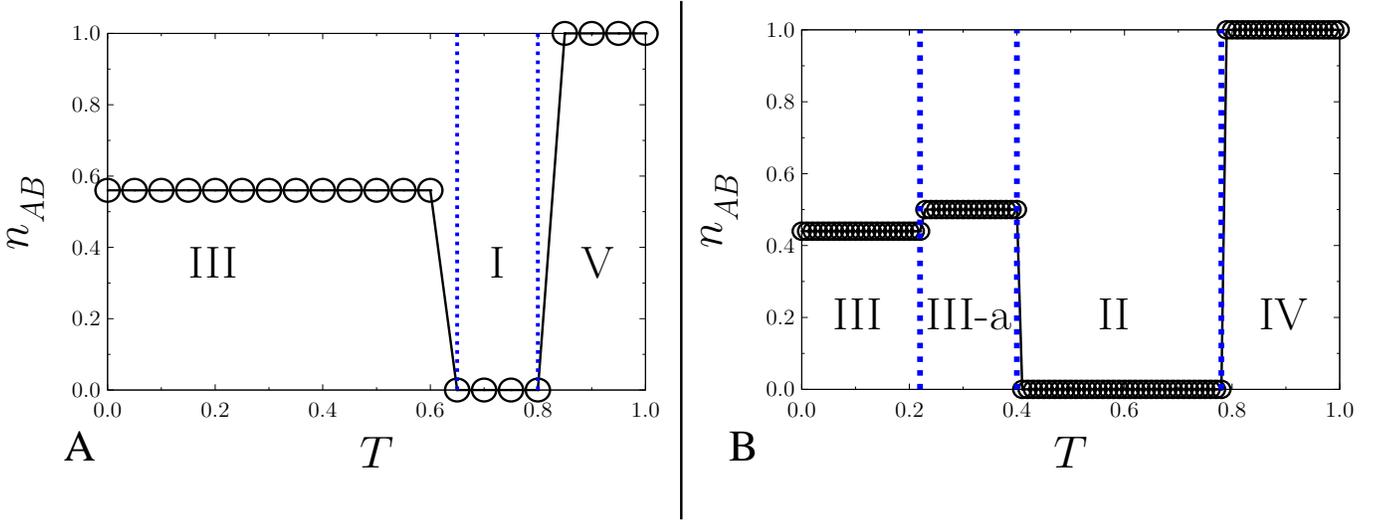}
 \caption{Panel (A): Bifurcation diagram among solutions III, I and  V of the average of $n_{AB}$ as function of T in the SCG
 for a fixed initial condition $x_{aa}=0.20$ and $x_{bb} = 0.20$. Panel (B): Bifurcation diagram among solutions III, III-a, II and IV of the average of $n_{AB}$ as
 function of T for a fixed initial condition $x_{aa}=0.78$ and $x_{bb} = 0.60$ where the risk parameter $b=3$ in the ACG.}
  \label{F7}
  \end{center}
 \end{figure}

We have shown in the previous section that the solution obtained for the SCG, and for a fixed initial condition, depends on the value of the threshold parameter $T$, so that by varying $T$ we find transitions among those solutions.
Examples of these transitions are shown in Figure \ref{F6}A to \ref{F6}C. These are bifurcation diagrams that give
the average of the fractions of inter-layer active links  $n_{AB}$ or $1- n_{AB}$ as a function of the threshold $T$.
These bifurcation diagrams describe transitions that occur, for threshold values of $T$, between solutions III to I,
III to V and I to V respectively
for different fixed values of the initial conditions. Each panel shows two examples.
We also find subsequent transitions among three solutions.
For instance, the bifurcation diagram   
Figure \ref{F7}A for the average of $n_{AB}$, illustrates a first transition between solution III and solution I, followed by a second transition between I and V  as  $T$ increases.
These results show the effect of the skepticism on the collective behavior in a two-layer network. Tuning the level of skepticism from the limit value  $T=0$, where the population is extremely {\it herding}  to
the limit value  $T=1$ where the population is extremely {\it skeptical}, Figure \ref{F4} indicates that the system goes from a state of complete coexistence of strategies, Solution III for almost all initial conditions,  to states of anticoordination, Solutions IV and V, and states of global coordination, Solutions I and II.

%%%%%%%%%%%%%%%%%%%%%%%%%%%%%%%%%%%%%%%%%%%%%%%%%%%%%%%%%%%%%%%%%%%%%%%%%%%%%%%%%%%
%Figure 6
 %
 %%%%%%%%%%%%%%%%%%%%%%%%%%%%%%%%%%%%%%%%%%%%%%%%%%%%%%%%%%%%%%%%%%%%%%%%%%%%%%%%%%%%
% Figure 7
 %
 %%%%%%%%%%%%%%%%%%%%%%%%%%%%%%%%%%%%%%%%%%%%%%%%%%%%%%%%%%%%%%%%%%%%%%%%%%%%%%%%%%%%%%
 %
\subsection{Asymmetric Coordination Game}
As discussed before, Figure \ref{F2} shows for the ACG case a phase diagram, obtained from a mean field theoretical approach, indicating domains of existence of different asymptotic solutions
in the $b$ - $T$ parameter space. In comparison with the SCG case, the additional parameter $b$ allows for new transitions that occur for a threshold value of $b$ and fixed $T$, including transition to chimera states. Examples of these transitions are shown in Figures \ref{F6}D to \ref{F6}G. Figs \ref{F6}D  shows a transition between a state of anticoordination IV and a state of full coordination I and  Fig.\ref{F6}G shows a transition between a state of full coordination II and a state of anticoordination V, while Figs. \ref{F6}E and \ref{F6}F show transitions between a state of coordination II or dynamical coexistence III and a chimera state III-a. On the other hand,
Figure \ref{F7}B shows an example of subsequent transitions as $T$ increases for fixed $b=3$, with a first transition between a state of dynamical coexistence III and a chimera state III-a,
followed by a transition between III-a and  a state of full coordination II and a final transition between II and
a state of anticoordination IV.

\begin{figure}[]
 \begin{center}
    \includegraphics[width=1\linewidth,angle=0]{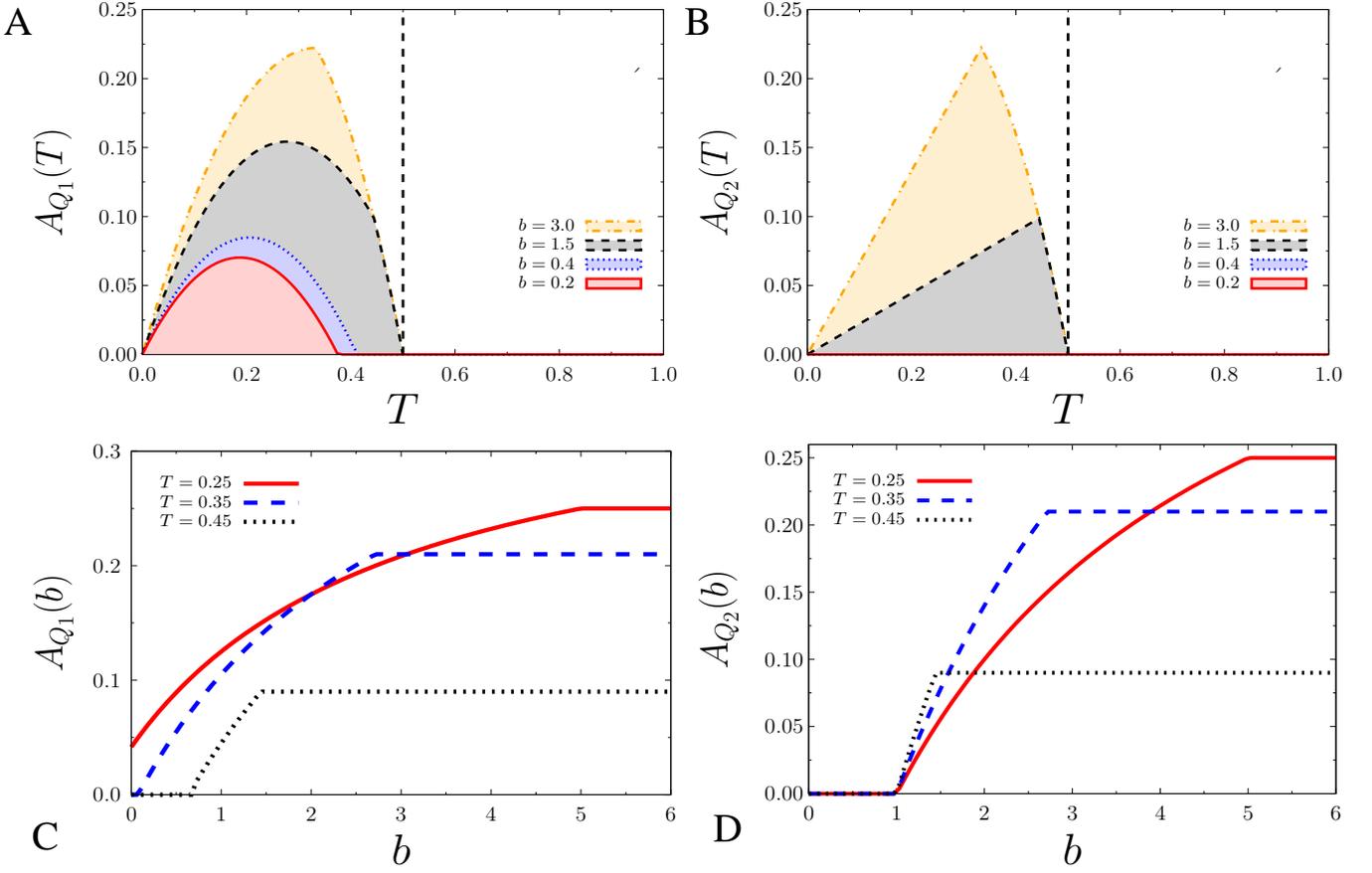}
 \end{center}
 \caption{Sizes of the basins of attraction of chimera states categorized by $Q_1$ and $Q_2$, denoted by $A_{Q_1}$  and $A_{Q_2}$, as function of T for different values of b (Panels A and B)
 and as function of b for different values of T (Panels C and D).}
  \label{F8}
 \end{figure}

A different form of bifurcation diagrams can be obtained by considering the area of the basin of attraction of a given solution in the parameter space of the initial conditions ${x_{aa}}^0$ and ${x_{bb}}^0$. Results for this area are indicated in Table \ref{rarea} for the chimera states $Q_1$ and $Q_2$ and the different zones of the $T$-$b$ parameter space of Fig.2. We recall that $Q_1$ are those solutions III-a and III-b reached from initial conditions such that ${x_{aa}}^0 + {x_{bb}}^0 <1$, while $Q_2$ are those obtained when ${x_{aa}}^0 + {x_{bb}}^0 >1$. The areas of the basins of attractions  $A_{Q_1}$, $A_{Q_2}$ for $Q_1$ and $Q_2$ respectively are plotted versus $T$ and $b$ in Figure \ref{F8}. Using $A_{Q_1}$, $A_{Q_2}$ as order parameters, these figures show bifurcation diagrams for the transition from existence $A_{Q}\neq0$ to non-existence $A_{Q}=0$ of a chimera state.

Figures \ref{F8}A and \ref{F8}B show a threshold value $T= 0.5$, so that chimera states exist for $T<0.5$ in agreement with
the phase diagram of Fig. 2. They also show that as $T$ increases the areas of the basin of attraction of chimera
states first increase until a certain value of $T$ and then they decrease to become zero for $T=0.5$.
In addition Figures \ref{F8}C and \ref{F8}D identify a threshold value of $b$ for the existence of chimera
states. For $Q_2$ chimeras this is fixed at $b=1$ independently of $T$, while for $Q_1$ it depends on $T$, with $Q_1$
chimera states existing for all values of $b$ and $T$ small enough. For both $Q_1$ and $Q_2$ we also identify a
characteristic $T$-dependent value of $b$ beyond which the area of the basin of attraction remains constant.

%%%%%%%%%%%%%%%%%%%%%%%%%%%%%%%%%%%%%%%%%%%%%%%%%%%%
% Table IV
\begin{table}[]
\centering
%\fontsize{6}{4}
\begin{tabular}{|l|l|l|l|l|}
\hline
zones                   			& ranges                  & $Q_1$ 		& $Q_2$ 		       \\ \hline
A        		& $ T < \frac{1+b}{3+b} < 0.5$	            & $2\left(\frac{1+b}{3+b} -T \right) T  $     		& 0         			 \\ \hline
B      			& $ \frac{1+b}{3+b} < T < 0.5 $	    & 0        		& 0       			\\ \hline
C 			&$ 1-  \frac{1+b}{3+b}< T <0.5$		& $2(1-2T)T$    		& $2(1-2T)T$        			       \\ \hline
D 			&$  T <1 -\frac{1+b}{3+b}<0.5 $        & $2\left(\frac{1+b}{3+b} - T \right) T  $     		& $2\left(2\frac{1+b}{3+b} -1 \right)T  $          \\ \hline
\end{tabular}
\caption{Areas of the basin of attraction of chimera states in the parameter space of initial conditions of $x_{aa}$ and $x_{bb}$  according to the zones described in Figure \ref{F2}. }
\label{rarea}
\end{table}
%
%%%%%%%%%%%%%%%%%%%%%%%%%%%%%%%%%%%%%%%%%%%%%%%%%%%%%%%%%%%
  % Figure 8

 %%%%%%%%%%%%%%%%%%%%%%%%%%%%%%%%%%%%%%%%%%%%%%%%%%%%%%%%%%%%%%%
More generally and on a qualitative basis, it follows from Figs. 4 and 5 that, for any fixed $b$,
the basins of attraction of solution III and chimeras states disappear as $T$ increases, so that and in the
limiting case of $T=1$, only Solutions I, II, IV and V can be reached by the system.
Another interesting limiting case is the one of the risk parameter $b  \rightarrow \infty$,
where it is extremely {\it risky} to play strategy $R$. It can be expected that in this limit solution I becomes preponderant. Indeed, we show in Fig. 9, as compared with Fig.5, that
the basin of attraction of solution I increases, solution II disappears and solution III and chimeras remain for fixed values of $T<0.5$.
As both $b$ and $T$ increase, solutions II, III, IV and V disappear and the basin of attraction of solution I increases. In the limit, $T=1$
and $b \rightarrow \infty$, solution I becomes the main solution in the system for almost every initial condition.
%%%%%%%%%%%%%%%%%%%%%%%%%%%%%%%%%%%%%%%%%%%%%%%%%%%%%%%%%%%%%%%%%%%%%%%%%%%%%%%%%%%%%
%Figura 9
\begin{center}
\begin{figure}[]
 \begin{center}
\includegraphics[width=1\linewidth,angle=0]{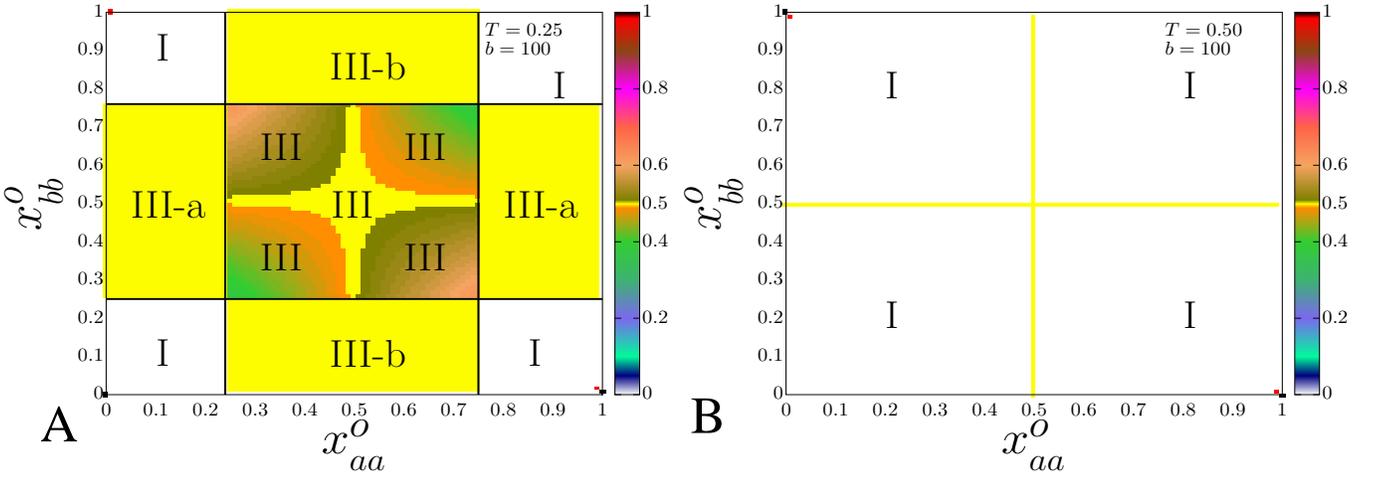}
 \end{center}
\caption{Plot in color scale of the fraction of active links between layer A and B in the asymptotic solution of the dynamics as a function of
   the initial density of $x_{aa}$ and $x_{bb}$ for the Asymmetric Coordination Game. The color scale defines the values
   of the fraction of actives links, $n_{AB}=1$ black color and $n_{AB}=0$
   white color. Asymptotic solutions are as indicated. Panel A: $b=100$, $T=0.25$, and panel B: $b=100$, $T=0.50$.}
\label{F9}
\end{figure}
\end{center}
%  %%%%%%%%%%%%%%%%%%%%%%%%%%%%%%%%%%%%%%%%%%%%%%%%%%%%%%%%%%%%%%%%%%%%%

 \section{Conclusions}

We have considered a model of evolutionary game of a population divided into two groups where individuals are searching to fulfil their social and strategic objectives.  The frame for this situation has been a multilayer network of two layers. Interactions within each layer aim to fulfil social objectives associated with learning dynamics, while interactions across layer consist in a coordination game, therefore involving strategic objectives. Our analysis, based on a mean-field theoretical approach and corroborated by numerical simulations of the model,  reveals the existence of collective behaviors commonly observed on nature but impossible to find on a single isolated  network analysis.
In our multilayer framework we find states different of those of global coordination or dynamical states of coexistence of the strategies. Namely,
in the multilayer coordination challenge, anticoordination and chimera states solution emerge. In the former
the dynamics of the system polarizes the population, with each layer coordinating in a different strategy. This can also happen in the asymmetric coordination game where the two strategies correspond to different Nash equilibria: the socially efficient or Pareto dominant, and the risk dominant equilibrium.
In the chimera states one layer coordinates in the risk dominant equilibrium, while the second remains disordered,
that is with coexistence of the two strategies. This coexistence can be time independent or in the form of periodic solutions.

In connection with the standard notion of chimera states in two populations of
dynamical oscillators having global or long range interactions \cite{as,Tinsley,Laing,Omel}, we note that in our evolutionary game theory framework we also have the basic ingredients of two nonlinear dynamical systems which are globally coupled. In chimera states of coupled oscillators, one population is in a coherent state and coexists
with the other population in an incoherent state. In our social analogue of the chimera state we have interpreted the coordination states in one layer as a coherent or ordered state,
while we identify the incoherent state with the layer that exhibits coexistence of the two strategies.
In most cases this coexistence is of dynamical nature, being the disordered layer in an active dynamical state of oscillation between the two possible strategies. Our model only incorporates two possible individual states of the agents, but we envisage that in social models including more individual states or strategies, such as those in reference (\cite{gac1,gac2}), the disordered state would show a richer dynamical behavior,
since the individual elements can dynamically visit a large number of possible states. In this case the disordered or incoherent population would have a dynamical behavior similar to those found in populations of dynamical oscillators.

We observe chimera states only for the asymmetric coordination game where the coherent state reached is in the
socially least desired coordination state. For herding individuals, the presence of a degree of risk in coordinating on the socially efficient
outcome has an effect on the emergence of chimera states. However, for skeptical individuals, the anticoordination states are present in both
symmetric and asymmetric coordination games. While the presence of two layers in the network is a consequence of the
type of interaction that individuals have inside and across the layers, the actual factors that play a key role for
the existence of chimera and anticoordination states are the level of skepticism and the existence of a risk parameter
on the coordination game.

In the context of coordination in social systems, our contribution brings a more realistic insight about the consequences of a collective behavior that makes a distinction between social and strategic objectives. This collective behavior may  lead herding societies to chimera states and skeptical societies to polarized states of anticoordination.

\section*{APPENDIX}

\subsection*{Mean-field approach: symmetric coordination game}

We present here the mean-field equations
for the time evolution of a system divided in two groups.
Our analysis reveals a highly nontrivial behavior of the learning and
coordinating process depending on the initial conditions of the system.

Let $x_{aa}$ be the fraction of individuals playing $R$ in
layer $A$ and $x_{bb}$ be the fraction of individuals playing $R$ in layer $B$.
In the limit of infinite population size ($N_A = N_B \rightarrow \infty$), and according to the levels of satisfaction described in Section 2 for the symmetric coordination
case,
the time derivative $\dot{x}_{aa}$ is given by:
\[
\begin{array}{rl}
\dot{x}_{aa}= & -x_{aa} \Theta(1 - x_{aa} - T) \Theta^{*}(1/2 - x_{bb}) H(x_{bb} =1) \\%  su,es
&  -x_{aa}[1- \Theta(1- x_{aa} -T )] H(x_{bb} < 1)\Theta^*(1/2 - x_{bb}) \Theta^*(1-x_{aa}) \\ %  ss,eu
& -x_{aa} \Theta(1 - x_{aa} - T) H(x_{bb} <1) \\ %  su,eu
& +(1-x_{aa}) \Theta( x_{aa} - T)  \Theta^*( x_{bb}-1/2) H(x_{bb}=0 ) \\ % su,es
& +(1-x_{aa}) [1- \Theta( x_{aa}-T)  ]H(x_{bb} >0) \Theta^*( x_{bb}-1/2)  \Theta^*(x_{aa}) \\ % ss,eu
& +(1-x_{aa}) \Theta( x_{aa}-T) H(x_{bb}> 0) %  su,eu
\end{array}
\]
 where $\Theta(z) = 1$ if $z \ge 0$ and $\Theta^{*} (z) = 1$ if $z > 0$, otherwise $\Theta(z) = 0$, and $H(z)= 1$ if $z$ is true, otherwise is 0.
Since the inequalities $1/2 >  x_{bb}$ and $ x_{bb} = 1$ cannot hold simultaneously, and neither $ x_{bb} > 1/2$ and $x_{bb} = 0$, the previous equation can be shortened as:
 \[
\begin{array}{rl}
\dot{x}_{aa}= &  -x_{aa}[1- \Theta(1- x_{aa} -T )]H(x_{bb} < 1) \Theta^*(1/2 - x_{bb}) \Theta^*(1-x_{aa})\\ %ss,eu
& -x_{aa} \Theta(1 - x_{aa} - T) H(x_{bb} <1)\\ %su,eu
& +(1-x_{aa}) [1- \Theta( x_{aa}-T) ]H(x_{bb} >0) \Theta^*( x_{bb}-1/2) \Theta^*(x_{aa})\\ % ss,eu
& +(1-x_{aa}) \Theta( x_{aa}-T)H(x_{bb}> 0 ) %  su,eu
\end{array}
\]
As mentioned before,  the equation is derived considering the different levels of satisfaction of agents playing $R$ or $L$  in group $A$.
Next, in order to simplify the analysis  we consider two cases  depending on whether $T$ is bigger than $1/2$. 
Table  \ref{mean2}
summarizes the results for $ \dot{x}_{aa}$ for the different values of $x_{aa}$ and $x_{bb}$ when we distinguish the two cases, $T> 1/2$ and $T<1/2$.
%
%table 5
\begin{table}[]
%\begin{center}
\begin{tabular}{|c|c|c|}
%\hline
value for $ \bf x_{aa} $  & $  \bf \dot{x}_{aa}$  &  value for $ \bf x_{bb}$  \\ \hline
0  & 0 & $ [0,1]$  	\\ \hline
$(0,1-T)$ &
$
\begin{array}{rl}
 -x_{aa} \\
 1 -2 x_{aa} \\
 1 - x_{aa}
\end{array}
$
&
$
 \begin{array}{rl}
[0,1/2]\\
 (1/2, 1)\\
1
\end{array}
$   \\  \hline
$(1-T,T)$ &
$
\begin{array}{rl}
 -x_{aa} \\
0 \\
 1 - x_{aa}
\end{array}
$
&
$
 \begin{array}{rl}
[0, 1/2)\\
1/2\\
 (1/2,1]
\end{array}
$  \\  \hline
$(T,1)$ &
$
\begin{array}{rl}
 -x_{aa} \\
 1 -2 x_{aa} \\
 1 - x_{aa}
\end{array}
$
&
$
 \begin{array}{rl}
 0\\
(0 , 1/2)\\
$ [1/2,1]$
\end{array}
$   \\  \hline
$1$ &
0
&
$[0,1]$   \\  \hline
\end{tabular}
%%%%%%%%%%%%%%%%%%%%%%%%%%%%%%%%%%%%%%%%%%%%%%%%%%
\begin{tabular}{c}
{}
\end{tabular}
\begin{tabular}{|c|c|c|}
%\hline
value for  $\bf x_{aa} $  & $\bf  \dot{x}_{aa}$  & value for $ \bf x_{bb}$  \\ \hline
0  & 0 & $ [0,1]$  	\\ \hline
$(0,T)$ &
$
\begin{array}{rl}
 -x_{aa} \\
 1 -2 x_{aa}  \\
 1 - x_{aa}
\end{array}
$
&
$
 \begin{array}{rl}
[0,1/2]\\
 (1/2, 1)\\
1
\end{array}
$   \\  \hline
$(T,1-T)$ &
$
\begin{array}{rl}
 -x_{aa} \\
 1 -2 x_{aa}  \\
 1 - x_{aa}
\end{array}
$
&
$
 \begin{array}{rl}
0\\
(0,1)\\
 1
\end{array}
$  \\  \hline
$(1-T,1)$ &
$
\begin{array}{rl}
 -x_{aa} \\
 1 -2 x_{aa}  \\
 1 - x_{aa}
\end{array}
$
&
$
 \begin{array}{rl}
 0\\
(0 , 1/2)\\
$ [1/2,1]$
\end{array}
$   \\  \hline
$1$ &
0
&
$[0,1]$   \\  \hline
\end{tabular}
\caption{Results of $ \dot{x}_{aa}$ for the different values of $x_{aa}$ and $x_{bb}$ in the SCG. $T > 1/2$ (left) and  $T < 1/2$ (right).}
\label{mean2}
%\end{center}
\end{table}%
Carrying out the same analysis for $\dot{x}_{bb}$ we can easily check that
the solutions are as follows:
\begin{itemize}
 \item Time independent coordination or anticoordination solutions:
 \begin{enumerate}
\item $ x_{aa} = x_{bb}= 0$. Solution I,
\item $ x_{aa}= x_{bb}=1$. Solution II,
\item $ x_{aa}=0$, $ x_{bb}= 1$. Solution IV,
\item $ x_{aa}=1$, $ x_{bb}=0$. Solution V.
\end{enumerate}
 \item Time independent coexistence solution: $x_{aa} = x_{bb}=1/2$. Solution III(1).
 \item A family of periodic solutions of coexistence: $ x_{aa}$ oscillates between $s$ and $1-s$ and $ x_{bb}$ between $r$ and $1-r$ for some  $0<s,r<1$. Solutions III(2).
\end{itemize}
Figure \ref{F4}
shows results of numerical simulations of our individual based model for different parameter values and different initial conditions ${x_{aa}}^0$ and ${x_{bb}}^0$ for $x_{aa}$ and
$x_{bb}$ respectively. These results identify the basins of attraction of the different solutions I, II, III, IV and V.
%%%%%%%%%%%%%%%%%%%%%%%%%%%%%%%%%%%%%%%%%%%%%%%%%%%%%%%%%%%%%%%%%%%%%%%%%
\subsection*{Mean-field approach: asymmetric coordination game}
By the same arguments than for the symmetric coordination case, and in the limit of infinite
 population size ($N_A = N_B \rightarrow \infty$),
the time derivative $\dot{x}_{aa}$ is given by:
%
%%%%%%%%%%%%%%%%%%%%%%%%%%%%%%%%%%%%%%%%%%%%%%%%%%%%%%%%%%%%%%%%%%%
\[
\begin{array}{rl}
\dot{x}_{aa}=
&  -x_{aa}[1- \Theta(1- x_{aa} -T )] H(x_{bb} < 1) \Theta^*(\frac{1+b}{3+b}  - x_{bb})\Theta^*(1-x_{aa}) \\%  ss, eu
& -x_{aa} \Theta(1 - x_{aa} - T) H(x_{bb} <1) \\ %  su, eu
& +(1-x_{aa}) [1- \Theta( x_{aa}-T) ]H(x_{bb} >0)\Theta^*(x_{bb} - \frac{1+b}{3+b})   \Theta^*(x_{aa}) \\ %  ss, eu
& +(1-x_{aa}) \Theta( x_{aa}-T) H(x_{bb}> 0 ) \\% su, eu
\end{array}
\]

The term $ \frac{1+b}{3+b}$ arises when the partially unsatisfied player with strategy $R$
will change to strategy $L$ if the players with the strategy $L$ gain a higher payoff than her.
This happens when the  fraction of players playing $R$ in the other layer is not larger than $ \frac{1+b}{3+b}$.
Table \ref{mean4} summarizes the results for $ \dot{x}_{aa}$ in the two cases, $T> 1/2$ and $T<1/2$. As in the symmetric coordination case, the equations display
the time independent coordinationn or anticoordination solutions: $x_{aa}=x_{bb}=0$ and $x_{aa}=x_{bb}=1$ and $x_{aa}$ and $x_{bb}$ equal to $0$ or $1$,  the time independent coexistence solution
$x_{aa} = x_{bb} = 1/2$ for $T<1/2$ and the family of periodic solutions of coexistence with
$0<x_{aa} < 1$ and $0< x_{bb} < 1$. Additionally,
we find three new types of  solutions:
\begin{enumerate}
\item For $T>1/2$.: a time independent coexistence solution: $x_{aa} = x_{bb} = \frac{1+b}{3+b}$. Solution III(1).
\item For $T<1/2$: Two chimera solutions with one dynamically disordered layer:
\begin{itemize}
\item   $ x_{aa} = 0$, and a periodic solution for $x_{bb}$. Solution III-a. 
\item   $x_{bb} = 0$ and a periodic solution for $x_{aa}$. Solution III-b. 
\end{itemize}
\item For $T<1/2$: Two chimera solutions with one time independent disordered layer:

$x_{aa} = 0$, $x_{bb} = 0.5$  or $x_{aa} = 0.5$, $x_{bb} = 0$. Solutions III-a or III-b.
\end{enumerate}
%
% table 6
\begin{table}[]
%\begin{center}
\begin{tabular}{|c|c|c|}
%\hline
value for $ \bf x_{aa} $  & $  \bf \dot{x}_{aa}$  &  value for $ \bf x_{bb}$  \\ \hline
0  & 0 & $ [0,1]$  	\\ \hline
$(0,1-T)$ &
$
\begin{array}{rl}
 -x_{aa} \\
 1 -2x_{aa}  \\
 1 - x_{aa}
\end{array}
$
&
$
 \begin{array}{rl}
 [0, \frac{1+b}{3+b}] \\
 (\frac{1+b}{3+b}, 1)\\
1
\end{array}
$   \\  \hline
$(1-T,T)$ &
$
\begin{array}{rl}
 -x_{aa} \\
0 \\
 1 - x_{aa}
\end{array}
$
&
$
 \begin{array}{rl}
[0, \frac{1+b}{3+b})\\
\frac{1+b}{3+b} \\
 (\frac{1+b}{3+b},1]
\end{array}
$  \\  \hline
$(T,1)$ &
$
\begin{array}{rl}
1-2x_{aa} \\
 1 - x_{aa}
\end{array}
$
&
$
 \begin{array}{rl}
[0, \frac{1+b}{3+b})\\
 \left[\frac{1+b}{3+b},1\right]
\end{array}
$
  \\  \hline
$1$ &
0
&
$[0,1]$   \\  \hline
\end{tabular}
%%%%%%%%%%%%%%%%%%%%%%%%%%%%%%%%%%%%%%%%%%%%%%%%%
\begin{tabular}{c}
{\quad}
\end{tabular}
\begin{tabular}{|c|c|c|}
%\hline
value for  $\bf x_{aa} $  & $\bf  \dot{x}_{aa}$  & value for $ \bf x_{bb}$  \\ \hline
0  & 0 & $ [0,1]$  	\\ \hline
$(0,T)$ &
$
\begin{array}{rl}
 -x_{aa} \\
 1 -2 x_{aa} \\
 1-x_{aa}
\end{array}
$
&
$
 \begin{array}{rl}
[0,\frac{1+b}{3+b}]\\
 (\frac{1+b}{3+b}, 1)\\
 1
\end{array}
$   \\  \hline
$(T,1-T)$ &
$
\begin{array}{rl}
 1 -2 x_{aa}  \\
 1 - x_{aa}
\end{array}
$
&
$
 \begin{array}{rl}
[0,1)\\
1
\end{array}
$  \\  \hline
$(1-T,1)$ &
$
\begin{array}{rl}
 1 -2 x_{aa} \\
 1 - x_{aa}
\end{array}
$
&
$
 \begin{array}{rl}
[0 , \frac{1+b}{3+b})\\
 \left[\frac{1+b}{3+b},1\right]
\end{array}
$   \\  \hline
$1$ &
0
&
$[0,1]$   \\  \hline
\end{tabular}
\caption{Results for $ \dot{x}_{aa}$ for the different values of $x_{aa}$ and $x_{bb}$ in the ACG. For $T > 1/2$, left side, and  $T < 1/2$, right side.}
\label{mean4}
%\end{center}
\end{table}%

\section*{Acknowledgments}
Hayd\'ee Lugo acknowledges financial support from Ministerio de Econom\'ia y Competitividad (Spain) under Project No. ECO2016-75992-P. Maxi San Miguel acknowledges financial support from Agencia Estatal de Investigaci\'on (AEI, Spain) and Fondo Europeo de Desarrollo Regional under Project ESoTECoS Grant No. FIS2015-63628-C2-2-R (AEI/FEDER,UE) and the Spanish State Research Agency, through the Mar{\'\i}a de Maeztu Program for Units of Excellence in R\&D (MDM-2017-0711).


\begin{thebibliography}{}

\bibitem{kb}  Kuramoto Y, Battogtokh D. Coexistence of Coherence and Incoherence in Nonlocally Coupled Phase Oscillators. {\em Nonlinear
Phenom. Complex Syst.}  (2002) {\bf 5}, 380-385.
%
\bibitem{as} Abrams DM,  Strogatz SH. Chimera States for Coupled Oscillators.
{\em Phys. Rev. Lett. }(2004) 93, 174102.
%
\bibitem{mp} Martens EA, Panaggio MJ, Abrams DM. Basins of attraction for chimera states.
{\em New J. Phys.}  (2016) {\bf18}, 022002.
%
\bibitem{rb} Rakshit S, Bera BK, Perc M, Ghosh D. Basin stability for chimera states.
{\em Scientific Reports} (2017) {\bf 7}, 2412.
%
\bibitem{so} SudaY, Okuda K. Persistent chimera states in nonlocally coupled phase oscillators.
 {\em Phys. Rev. E} (2015) {\bf 92}, 060901(R).
%
\bibitem{pa} Panaggio MJ, Abrams DM. Chimera states: Coexistence of coherence and incoherence in networks of coupled oscillators.
{\em Nonlinearity}(2015) {\bf 28}, R67-R87.
%
\bibitem{kbg} Kundu S, Majhi S, Bera BK, Ghosh D, Lakshmanan, M. Chimera states in two-dimensional networks of locally coupled oscillators.
{\em Phys Rev. E } (2018) {\bf 97}, 022201.
%
\bibitem{gac1} Gonz\'alez-Avella JC, Cosenza MG, San Miguel M. A model for cross-cultural reciprocal interactions through mass media
{\em PLOS ONE} (2012) {\bf 7}, e51035.
%
\bibitem{gac2} Gonz\'alez-Avella JC, Cosenza MG, San Miguel M. Localized coherence in two interacting populations of social agents.
{\em Physica A} (2014) {\bf 399}, 24-30. doi:10.1016/j.physa.2013.12.035.
%
\bibitem{a} Axelrod R, The Dissemination of Culture: A Model with Local Convergence and Global Polarization.
{\em J. Conflict Res}. (1997) {\bf 41},2,  203-226.
%
\bibitem{dna} Deffuant D, Neau F, Amblard G. Weisbuch. Mixing beliefs among interacting agents
{\em Advances in Complex Systems} (2000){\bf 3}, 87-98
%
\bibitem{redondo} Gonz\'alez-Avella JC, Egu\'iluz VM, Marsili M, Vega-Redondo F, San Miguel M. Threshold learning dynamics in social networks
{\em PLoS ONE} (2011){\bf 6(5)},e20207
%
\bibitem{e} Ellison G. Learning, local interaction, and coordination.
{\em Econometrica} (1993) {\bf 61},5,  1047-1071. doi: 10.2307/2951493
%
\bibitem{vr} Vega-Redondo, F. Economics and the Theory of Games; Cambridge University Press: Cambridge, UK (2003)
%
\bibitem{gh} Goeree JK, Holt CA.  Coordination games. In The Encyclopedia of cognitive science 2, 204-208,
ed. Nature Publishing Group, McMillan (2002).
%
\bibitem{lsm} Lugo H, San Miguel M. Learning and coordinating in a multilayer network.
{\em Scientifc Reports} (2015) {\bf 5}. doi:10.1038/srep07776
%
\bibitem{hs} Harsanyi J, Selten R. A General Theory of Equilibrium Selection in Games; The MIT Press:
Cambridge, MA, USA. (1988).
%
\bibitem{w} Weidenholzer S.  Coordination Games and Local Interactions: A Survey of the Game Theoretic Literature.
{\em  Games} (2010){\bf 1}(4), 551-585.
%

\bibitem{glsm} Gonz\'alez-Avella JC, Lugo H, San Miguel M. Coordination in a skeptical two-group population.
{\em J Econ Interact Coord }(2018). https://doi.org/10.1007/s11403-018-0223-x
%

\bibitem{Tinsley} Tinsley M. R, Nkomo S, Showalter K, Chimera and phase-cluster states in populations of coupled chemical oscillators \em {Nature Phys}. {\bf8} (2012).

\bibitem{Laing} Laing C. R, Chimeras in networks of planar oscillators. \em {Phys. Rev. E} {\bf81} (2010) 066221. https://doi.org/10.1103/PhysRevE.81.066221

\bibitem{Omel} Omelchenko I, Maistrenko Y, Hovel P, Scholl E, Loss of coherence in dynamical networks: spatial chaos and chimera states. Phys. Rev. Lett. 106 (2011) 234102

\end{thebibliography}
\end{document}